\documentclass[aps,preprint,showpacs, showkeys]{revtex4}%
\usepackage{amsfonts}
\usepackage{amsmath}
\usepackage{amssymb}
\usepackage{graphicx}%
\providecommand{\U}[1]{\protect \rule{.1in}{.1in}}

\begin{document}
\title[ ]{ Vacuum Stability of the wrong sign $(-\phi^{6})$ Scalar Field Theory }
\author{Abouzeid. M. Shalaby\thanks{E-mail:amshalab@ mans.edu.eg}}
\affiliation{Physics Department, Faculty of Science, Mansoura University, Egypt.}

\begin{abstract}
We apply the effective potential method to study the vacuum stability of the
bounded from above $(-\phi^{6})$ (unstable) quantum field potential. The
stability ($\partial E/\partial b=0)$ and the mass renormalization
($\partial^{2} E/\partial b^{2}=M^{2})$ conditions force the effective
potential of this theory to be bounded from below (stable). Since bounded from
below potentials are always associated with localized wave functions, the
algorithm we use replaces the boundary condition applied to the wave functions
in the complex contour method by two stability conditions on the effective
potential obtained. To test the validity of our calculations, we show that our
variational predictions can reproduce exactly the results in the literature
for the $\mathcal{PT}$-symmetric $\phi^{4}$ theory. We then extend the
applications of the algorithm to the unstudied stability problem of the
bounded from above $(-\phi^{6})$ scalar field theory where classical analysis
prohibits the existence of a stable spectrum. Concerning this, we calculated
the effective potential up to first order in the couplings in $d$ space-time
dimensions. We find that a Hermitian effective theory is instable while a
non-Hermitian but $\mathcal{PT}$-symmetric effective theory characterized by a
pure imaginary vacuum condensate is stable (bounded from below) which is
against the classical predictions of the instability of the theory. We assert
that the work presented here represents the first calculations that advocates
the stability of the $(-\phi^{6})$ scalar potential.

\end{abstract}
\maketitle

\section{Introduction}

\label{int}

The very active research area of the non-Hermitian theories with real spectra
\cite{bender,aboebt,ghost,aboeff,bendvs,spect,nonr, spect1,ghost1,ghost2} may
offer solutions to current existing problems in our understanding of nature.
Among the very large number of non-Hermitian theories investigated, theories
with bounded from above potentials deserve more interest than what offered in
the literature. In their quantum field versions, such theories possess the
very important asymptotic freedom property
\cite{aboebt,Symanzik,bendf,Frieder}. To shed light on the importance of this
property, one has to mention that in the past, to have such interesting
property, physicists had to resort to a somehow complicated model which merges
group theory to field theory with the number of colors to be equal to or
greater than three (quantum chromodynamics). Now and after the discovery of
possible physical acceptability of some of the non-Hermitian theories, one can
get the important asymptotic freedom property from just colorless, one
component and scalar field theories.

Apart from the above mentioned benefits that can be obtained from the
employment of the non-Hermitian theories in our modeling of natural events, a
big problem was thought to exist in dealing with such theories. In fact, for
physical amplitude calculations in the non-Hermitian theories, the metric
operator formulations are indispensable. However, the suggested regimes for
metric operator calculations in the literature turns the theory divergent even
at the quantum mechanical level of study \cite{bendvs}. For Higher dimensions,
the degree of divergences will be even higher and the calculation of the
metric operator becomes complicated and even hard to get in a closed form for
some perturbative calculations \cite{bendmet}. However, Jones and Rivers
showed that in case of metric operator of gauge form, one can get physical
amplitudes from path integrals within the non-Hermitian theory \cite{jonesqds}%
. Moreover, it has been shown that the effective field approach does know
about the metric \cite{jonesgr1,jonesgr2} and since effective field approach
can be easily extended to the important quantum field case
\cite{aboeff,aboebt}, one may not worry about the metric any more.

The complex contour method followed in the literature, apart from its success,
is hard to follow specially for the study of non-Hermitian quantum field
theories. In this approach, one finds a contour in the complex $\ x$-plane
($x$ is equivalent to the field variable in our work) on which the wave
functions are localized in the sense that a wave function vanishes as
$\left \vert x\right \vert \rightarrow0$. To avoid the problem of finding a
contour of such characteristics, one can think in taking the imaginary part of
$x$ as a variational parameter that can be adapted in such away that secures
the localization of wave functions. However, the process is similar to a field
shift which is a well known method in the literature through which one can
obtain the corresponding effective potential. The challenge now is to map the
boundary condition on wave functions on a complex contour to a condition on
the effective potential. In fact, our experience in solving quantum mechanical
problems tells us that bounded from below potentials results in localized wave
function. Accordingly, one can replace localized wave function boundary
condition in the complex contour method by constraining the parameters in the
effective potential in such away that turns it bounded from below (stable).

Although one can follow path integral to calculate the effective potential, in
this work we follow the canonical quantization method
\cite{efbook,Peskin,Coleman,normal2,Canon1,changcac,mag} to calculate the
effective potential. In fact, in the context of non-Hermitian theories, it
would be more suitable to follow the Canonical quantization method as the
Hamiltonian operator can reflect the non-Hermiticity of a theory more clearer
than dealing with classical functions in the path integral formalism. The
organization of the paper is as follows. In section \ref{form}, we introduce
the formulation of the effective potential within the canonical quantization
method. In section \ref{1d}, we study the quantum mechanical case ($0+1$
dimensions). In section \ref{high}, we study the the wrong sign $(-\phi^{6})$
scalar field theory in $1+1$ and $2+1$ space-time dimensions. Also, in section
\ref{comp}, we test our results by presenting comparisons with those available
in the literature while conclusion follows in section \ref{conc}.

\section{ The Calculation of the effective potential}

\label{form}

To start, consider the quantum field Hamiltonian density of the form;%
\begin{equation}
H=\frac{1}{2}\left(  \triangledown \phi \right)  ^{2}+\frac{1}{2}\pi^{2}%
+\frac{1}{2}m^{2}\phi^{2}+g\phi^{4}+h\phi^{6}, \label{ham1}%
\end{equation}
where $m$ is the mass of the field $\phi$, $\pi$ is the conjugated momentum
field while $g$ and $h$ are coupling constants. The mean field approach is
lunched by the application of the canonical transformation $\phi=\psi+B$ and
$\pi=\Pi$. Here, $B$ is a constant called the vacuum condensate and
$\Pi=\overset{\cdot}{\psi}$. The philosophy behind the field shift is to
account in variational manner for the imaginary part of the complex contour
method followed in the literature \cite{jones,bender}. This field shift will
lead to an effective potential on which one may apply constraints that is
equivalent to the boundary condition on a complex contour. In the literature,
we get used to have localized wave functions associated with bounded from
below potentials in quantum mechanical problems. Accordingly, one can
constrain the effective potential to be bounded from below and don't exclude
values of the parameters that turn the theory non-Hermitian. Equivalently, we
map the condition $\chi \left(  x\right)  \rightarrow0$ as $\left \vert
x\right \vert \rightarrow \infty$ ( $\chi \left(  x\right)  $ is the wave
function) in the complex contour algorithm to the constraint on the effective
potential that enforces it to be bounded from below. In this way we obtain a
more practical algorithm to do calculations even in complicated non-Hermitian
field theories for which one may fail to follow the complex contour algorithm.

Plugging the above transformations into the Hamiltonian model in
Eq.(\ref{ham1}) to get an equivalent effective form as;%
\begin{align}
H  &  =\frac{1}{2}\left(  \triangledown \psi \right)  ^{2}+\frac{1}{2}\Pi
^{2}+\frac{1}{2}M^{2}\psi^{2}+\left(  15hB^{4}+6gB^{2}+\frac{1}{2}m^{2}%
-\frac{1}{2}M^{2}\right)  \psi^{2}\nonumber \\
&  +\left(  20hB^{3}+4gB\right)  \allowbreak \psi^{3}+\left(  15hB^{2}%
+g\right)  \psi^{4}+6Bh\psi^{5}+h\psi^{6}\label{effh}\\
&  +\left(  6hB^{5}+4gB^{3}+Bm^{2}\right)  \allowbreak \psi+\left(
hB^{6}+gB^{4}+\frac{1}{2}B^{2}m^{2}\right)  .\nonumber
\end{align}
We have chosen to work with the mass parameter $M$ ( which collects all the
coefficients of $\psi^{2}$) of the field $\psi$ \ and consider all terms
except the kinetic term $(\frac{1}{2}\left(  \triangledown \psi \right)
^{2}+\frac{1}{2}\Pi^{2}+\frac{1}{2}M^{2}\psi^{2})$ to constitute an
interaction Hamiltonian. The vacuum expectation value of the Hamiltonian
operator $\langle \Omega \left \vert H\right \vert \Omega \rangle$ is known as the
effective potential or vacuum energy, where $|\Omega \rangle$ represents the
vacuum state of the interacting theory. In the canonical quantization regime (
the first part of Ref. \cite{Peskin}), one can calculate the expectation value
of an operator in an interactive theory in terms of the free vacuum state
$|0\rangle$ via the employment of the time evolution operator ( Eq. (4.31) in
Ref. \cite{Peskin}). At the tree level, the vacuum energy is given by;%

\[
\langle \Omega \left \vert H\right \vert \Omega \rangle=E=hB^{6}+gB^{4}+\frac{1}%
{2}B^{2}m^{2}.
\]
In this form one realizes that;
\[
\frac{\partial E}{\partial B}=6hB^{5}+4gB^{3}+Bm^{2},
\]
which is exactly the coefficient of the linear term in the field $\psi$ in
Eq.(\ref{effh}). This term has to be dropped out if \ one seeks the stability
of the theory and this term will disappear order by order \cite{Peskin}.
Another realization is that $\frac{\partial^{2}E}{\partial B^{2}}$ is known to
be equal to $i\left(  \frac{i}{p^{2}-M^{2}}\right)  ^{-1}$, with $p$ as the
external momentum \cite{swanson}. Since $B$ is position independent, then the
external momentum is certainly zero. So one can employ the constraints;
\begin{equation}
\frac{\partial E}{\partial B}=0\text{ and }\frac{\partial^{2}E}{\partial
B^{2}}=M^{2}, \label{EBM}%
\end{equation}
on the parameters $B$ and $M$. If $M^{2}$ is to be chosen positive, this means
that the effective potential $E$ is bounded from below and thus stable. In our
work we will use the conditions in Eq. (\ref{EBM}) to mimic the localized wave
function boundary condition in the complex contour method in Ref.\cite{bender}.

\begin{figure}[ptbh]
\centering
\includegraphics{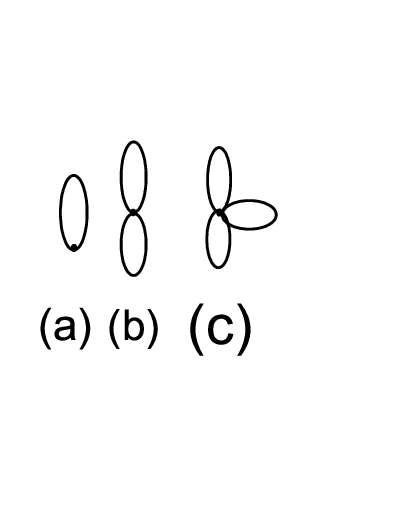} \caption{ Feynman diagrams contributing to the
first order radiative corrections to the vacuum energy of the $\phi^{4}%
+\phi^{6}$ theory. Diagram (a) generated from the full contraction of the
$\psi^{2}$ in the interaction Hamiltonian while diagrams (b) and (c) are
generated from the full contraction of the $\psi^{4}$ and $\psi^{6}$ terms in
the interaction Hamiltonian, respectively.}%
\label{feyn}%
\end{figure}

For the first order correction ( in the couplings) to the effective potential,
we consider the Feynman diagrams contributing to this order shown in
Fig.\ref{feyn}. \ Note that, one can split the vacuum expectation value of the
Hamiltonian as $\langle \Omega \left \vert H\right \vert \Omega \rangle
=\langle \Omega \left \vert H_{0}\right \vert \Omega \rangle+\langle \Omega
\left \vert H_{I}\right \vert \Omega \rangle$, where the Hamiltonian operator $H$
has been decomposed into the free Hamiltonian $H_{0}$ plus the interaction
Hamiltonian $H_{I}$. In fact, $H_{0}$ is diagonal with respect to the free
vacuum and thus can easily obtained to be;%
\[
\langle \Omega \left \vert H_{0}\right \vert \Omega \rangle=\frac{1}{2\left(
4\pi \right)  ^{\frac{d-1}{2}}}\left(  \frac{\Gamma \left(  -\frac{1}{2}%
-\frac{d-1}{2}\right)  }{\Gamma \left(  -\frac{1}{2}\right)  }\left(  \frac
{1}{M^{2}}\right)  ^{-\frac{1}{2}-\frac{d-1}{2}}\right)  ,
\]
and the the expectation value of the interaction Hamiltonian $\langle
\Omega \left \vert H_{I}\right \vert \Omega \rangle$ \ can be obtained
perturbatively. Note that the diagrams generated by the amplitude
$\langle \Omega \left \vert H_{I}\right \vert \Omega \rangle$ do not have external
legs (zero momentum) because the Hamiltonian itself is an integration over the
position space of the Hamiltonian density $H$. In other words, the diagrams
contributing to the amplitude $\langle \Omega \left \vert H_{I}\right \vert
\Omega \rangle$ are those generated by fully contracted internal lines.

Up to first order in the couplings, the fully contracted diagrams of the
theory under investigation are given in Fig. \ref{feyn}. Accordingly, one can
get the result of $\langle \Omega \left \vert H_{I}\right \vert \Omega \rangle$ for
$d$ space-time dimensions as;%
\begin{align}
E  &  =\frac{1}{2\left(  4\pi \right)  ^{\frac{d-1}{2}}}\left(  \frac
{\Gamma \left(  -\frac{1}{2}-\frac{d-1}{2}\right)  }{\Gamma \left(  -\frac{1}%
{2}\right)  }\left(  \frac{1}{M^{2}}\right)  ^{-\frac{1}{2}-\frac{d-1}{2}%
}\right)  +\frac{1}{2}B^{2}m^{2}+gB^{4}+hB^{6}+\left(  \frac{-i6!h}%
{-3!\times8i}\right)  \left(  \Delta \right)  ^{3}\nonumber \\
&  +\left(  \frac{-i4!\left(  15hB^{2}+g\right)  }{-i8}\right)  \left(
\Delta \right)  ^{2}+\left(  \frac{-i2\left(  15hB^{4}+6gB^{2}+\frac{1}{2}%
m^{2}-\frac{1}{2}M^{2}\right)  }{-i2}\right)  \left(  \Delta \right)  ,
\end{align}
$\allowbreak$

where%
\[
\Delta=\frac{1}{\left(  4\pi \right)  ^{\frac{d}{2}}}\frac{\Gamma \left(
1-\frac{d}{2}\right)  }{\Gamma \left(  1\right)  }\left(  \frac{1}{M^{2}%
}\right)  ^{1-\frac{d}{2}},
\]
and $\Gamma$ is the gamma function.This form of the vacuum energy is finite in
$0+1$ space-time dimensions but for higher dimensions divergences are existing
and need certain procedures to get red out of them.

\section{ the effective potential of the $\mathcal{PT}$-symmetric $(-\phi
^{6})_{0+1}$ theory}

\label{1d}

For $d=1$ (quantum mechanics), $E$ can be simplified as;
\begin{equation}
E=\frac{1}{2}M+\frac{1}{2}B^{2}m^{2}+gB^{4}+hB^{6}+\frac{3}{4}\frac{\left(
15hB^{2}+g\right)  }{M^{2}}+\frac{15}{8}\frac{h}{M^{3}}+\frac{1}{4M}\left(
30hB^{4}+12gB^{2}-M^{2}+m^{2}\right)  . \label{veff}%
\end{equation}
For this effective potential to be stable, one has to constrain the parameters
introduced in the calculations such that;%
\[
\frac{\partial E}{\partial B}=0.
\]
Let us first study the case of $g=0$ and $m=0$, where we get the result;%
\begin{equation}
\frac{3}{2}\frac{B}{M^{2}}h\left(  4B^{4}M^{2}+20B^{2}M+15\right)  =0.
\end{equation}
This equation has three different solutions of the form;%
\begin{align}
B  &  =0,\nonumber \\
B^{2}  &  =-\frac{1}{2M}\left(  \sqrt{10}+5\right)  ,\text{ }\label{bsol}\\
B^{2}  &  =\frac{1}{2M}\left(  \sqrt{10}-5\right)  .\nonumber
\end{align}
The $B=0$ solution is acceptable only for the bounded from below theory
(positive $h)$. In this case the theory is Hermitian and the vacuum is stable
as well. For the solutions $B^{2}=-\frac{1}{2M}\left(  \sqrt{10}+5\right)  $
and $B^{2}=\frac{1}{2M}\left(  \sqrt{10}-5\right)  $, $M$ is positive and thus
$B$ is imaginary. Accordingly, the Hamiltonian form in Eq.(\ref{effh}) is
non-Hermitian but $\mathcal{PT-}$symmetric as well and one then can claim that
the spectrum of the theory is real and stable for both broken symmetry
solutions in Eq.(\ref{bsol}) either $h$ positive or negative. In fact, the
story here is different and it is only the solution $B^{2}=-\frac{1}%
{2M}\left(  \sqrt{10}+5\right)  $ that is stable for the bounded from below
potential $\left(  +h\right)  $ while the solution $B^{2}=\frac{1}{2M}\left(
\sqrt{10}-5\right)  $ represents an unstable vacuum. In fact, for the solution
$B^{2}=-\frac{1}{2M}\left(  \sqrt{10}+5\right)  $ , the effective potential
has the form;%

\[
E=-\frac{1}{B^{2}\left(  136\sqrt{10}+440\right)  }\left(  \left(
64h+32\sqrt{10}h\right)  B^{8}+140\sqrt{10}+445\right)  \allowbreak,
\]
which shows that $E$ with $B$ real is $-E$ with $B$ imaginary. In fact, real
$B$ means that $M$ is negative which means the existence of ghost states
(negative kinetic energy).

\begin{figure}[ptb]
\centering
\includegraphics{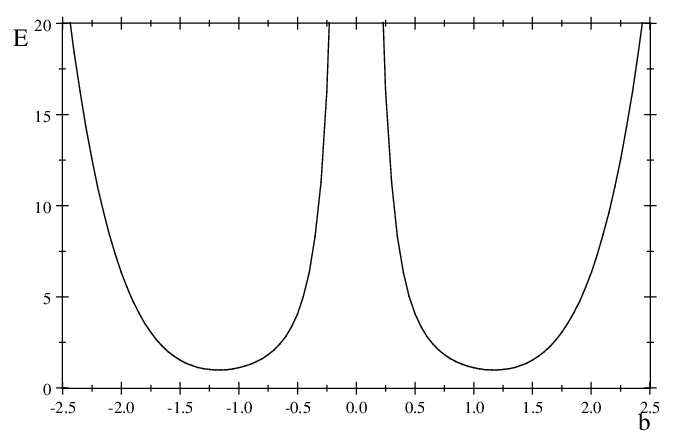} \caption{The ground state energy as a function of
the one-point function $B\equiv \left \langle 0|\phi|0\right \rangle \equiv ib $
measured in units of $i$ for $h=0.5$ for the solution $B^{2}=-\frac{1}%
{2M}\left(  \sqrt{10}+5\right)  $ of the $\mathcal{PT}$-symmetric $\phi^{6}$
potential.}%
\label{Ebhp5s}%
\end{figure}

\begin{figure}[ptbh]
\centering
\includegraphics{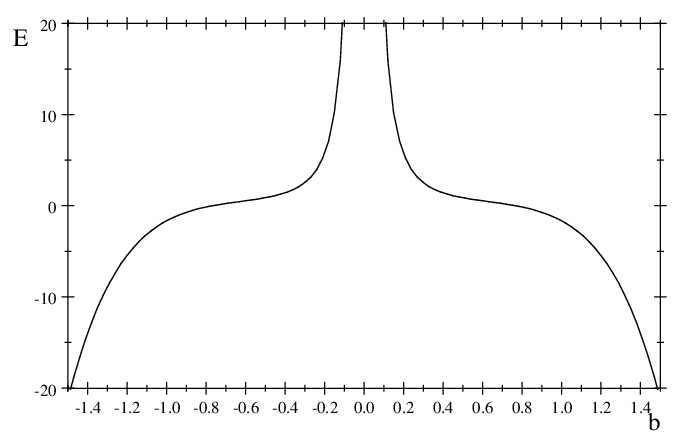} \caption{The ground state energy as a function
of the one-point function measured in units of $i$ for $h=0.5$ for the the
solution $B^{2}=\frac{1}{2M}\left(  \sqrt{10}-5\right)  $ of the
$\mathcal{PT}$-symmetric $\phi^{6}$ potential.}%
\label{Ebhp5is}%
\end{figure}

In this article we shall stick to the usual understanding of particles as they
have positive masses and stability exists from minimizing actions.
Consequently, $B$ is chosen imaginary and to investigate the stability of the
theory we plot the diagram in Figs.\ref{Ebhp5s} where one can realize that the
effective potential (vacuum energy ) is bounded from below for $h=+\frac{1}%
{2}$ for the solution $B^{2}=-\frac{1}{2M}\left(  \sqrt{10}+5\right)  $ . On
the other hand, the solution $B^{2}=\frac{1}{2M}\left(  \sqrt{10}-5\right)  $
results in an instable vacuum since the associated effective potential is
unbounded either from above or from below (Fig. \ref{Ebhp5is}). Again with the
solution $B^{2}=\frac{1}{2M}\left(  \sqrt{10}-5\right)  $, the effective
potential has the form;%

\[
E=\frac{1}{B^{2}\left(  136\sqrt{10}-440\right)  }\left(  \left(
64h-32\sqrt{10}h\right)  B^{8}-140\sqrt{10}+445\right)  \allowbreak,
\]
which for $B$ real has exactly an opposite sign to the imaginary $B$ result.

\begin{figure}[ptbh]
\centering
\includegraphics{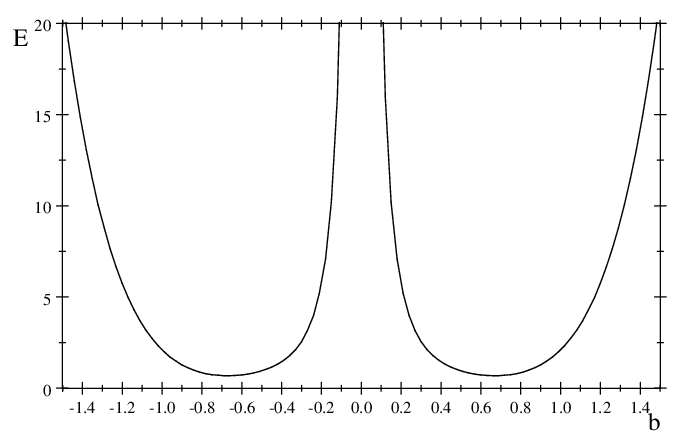} \caption{The ground state energy as a function of
the one-point function measured in units of $i$ for $h=-0.5$ for the the
solution $B^{2}=\frac{1}{2M}\left(  \sqrt{10}-5\right)  $ of the
$\mathcal{PT}$-symmetric $\phi^{6}$ potential.}%
\label{Ehnp5}%
\end{figure}

\begin{figure}[ptbh]
\centering
\includegraphics{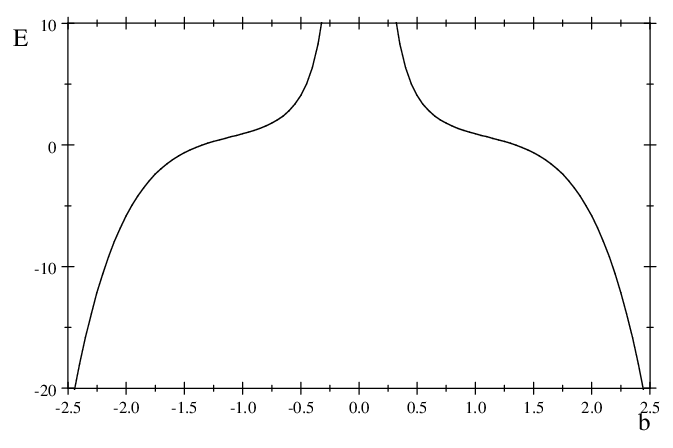} \caption{The ground state energy as a function
of the one-point function measured in units of $i$ for $h=-0.5$ for the
solution $B^{2}=-\frac{1}{2M}\left(  \sqrt{10}+5\right)  $ of the
$\mathcal{PT}$-symmetric $\phi^{6}$ potential.}%
\label{Ebhnp5is}%
\end{figure}

For an unstable classical potential (negative $h$ coupling), on the other
hand, the solution $B^{2}=\frac{1}{2M}\left(  \sqrt{10}-5\right)  $ results in
a stable effective potential as shown in Fig.\ref{Ehnp5} while the solution
$B^{2}=-\frac{1}{2M}\left(  \sqrt{10}+5\right)  $ is unstable
(Fig.\ref{Ebhnp5is}). These results are in fact very interesting since they
show that stability (like tunneling) can not always be argued in view of
classical analysis. A classically stable potential may or may not lead to a
stable quantized system. The reverse is also correct, a classically instable
potential can have stable as well as instable spectra.

\begin{figure}[ptbh]
\centering
\includegraphics{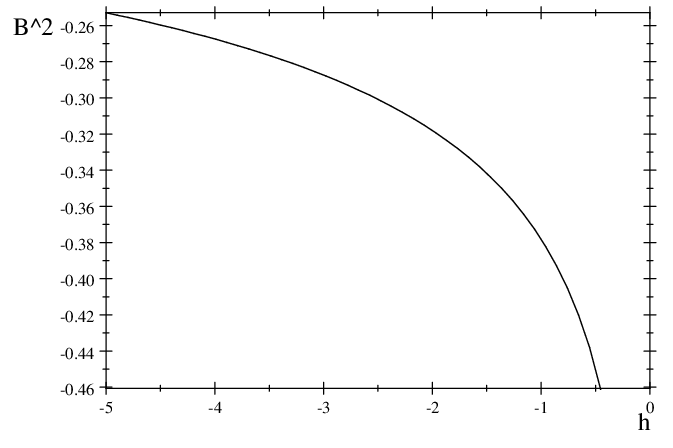}\caption{The one-point function squared versus the
coupling $h$ for the massless $\mathcal{PT}$-symmetric $(-\phi^{6})$ potential
in $0+1$ space-time dimensions.}%
\label{B2hn}%
\end{figure}

\begin{figure}[ptbh]
\centering
\includegraphics{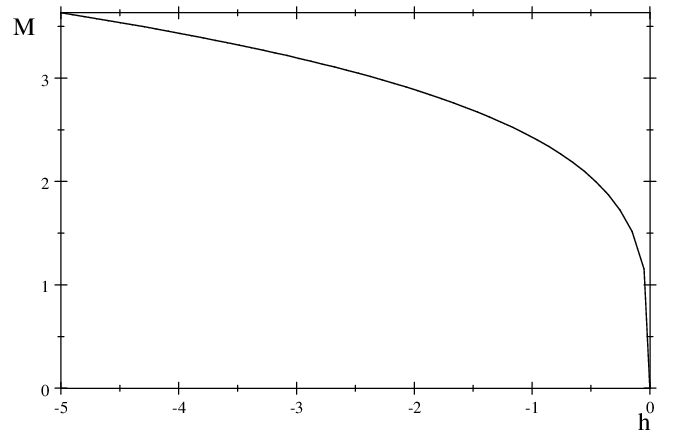} \caption{The effective mass of the field $\psi$
versus the coupling $h$ for the massless $\mathcal{PT}$-symmetric $(-\phi
^{6})$ potential in $0+1$ space-time dimensions.}%
\label{t-nh}%
\end{figure}

For the bounded from above potential (negative $h)$ the $B=0$ solution is
unstable. Accordingly, only the broken symmetry solution characterized by the
parameters $B^{2}=\frac{1}{2M}\left(  \sqrt{10}-5\right)  $ and $M=\sqrt[4]%
{-30\left(  \sqrt{10}-2\right)  h}$ is the only acceptable solution. For this
case, the behavior of the vacuum condensate as a function of the coupling $h$
is shown in Fig.\ref{B2hn} while the behavior of the effective mass $M$ is
presented in Fig.\ref{t-nh}.

The results above show that the algorithm we follow is reliable in studying
non-Hermitian theories. A big advantage of this algorithm is that it can be
extended easily to theories which have never been studied like the bounded
from above theories with many couplings.

In $d$ space-time dimensions, for the general case of massive theory as well
as for $g\neq0$, we get the results;
\begin{align}
M  &  =\frac{1}{\left(  15^{\frac{1}{-2+d}}\right)  }2^{\frac{d}{-2+d}}%
\pi^{\frac{d}{2\left(  -2+d\right)  }}\frac{\left(  -2g-10B^{2}h+\sqrt
{4g^{2}+40B^{4}h^{2}-10hm^{2}}\right)  ^{\frac{1}{-2+d}}}{\left(  h^{\frac
{1}{-2+d}}\right)  \left(  \left(  -\Gamma \left(  -\frac{1}{2}d\right)
d\right)  ^{\frac{1}{-2+d}}\right)  }\allowbreak,\nonumber \\
M  &  =\frac{1}{\left(  15^{\frac{1}{-2+d}}\right)  }2^{\frac{1}{-2+d}d}%
\pi^{\frac{1}{2\left(  -2+d\right)  }d}\frac{\left(  -2g-10B^{2}h-\sqrt
{4g^{2}+40B^{4}h^{2}-10hm^{2}}\right)  ^{\frac{1}{-2+d}}}{\left(  h^{\frac
{1}{-2+d}}\right)  \left(  \left(  -\Gamma \left(  -\frac{1}{2}d\right)
d\right)  ^{\frac{1}{-2+d}}\right)  }\allowbreak
\end{align}

\begin{figure}[ptbh]
\centering
\includegraphics{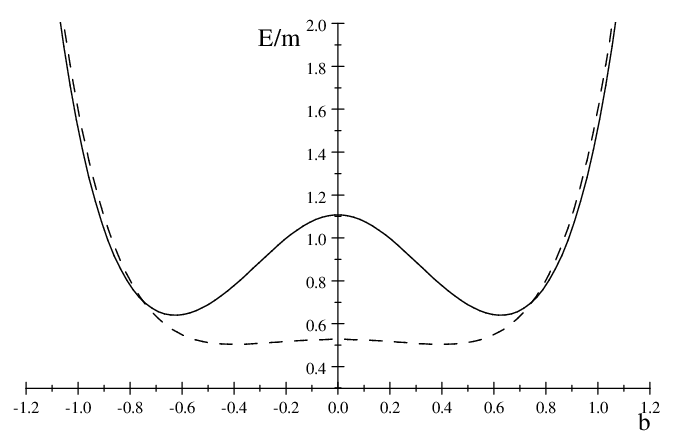} \caption{The $0+1$ space-time dimensions
effective potential as a function of the one-point function measured in units
of $i$ for $H=-\frac{1}{4}\mathtt{,} G=+\frac{1}{2}$ (dashed) and $H=-\frac
{1}{4}\mathtt{,} G=-\frac{1}{2}$ (solid) with the stable solution
$t=\frac{-15H}{2G-10b^{2}H+\sqrt{4G^{2}+40b^{4}H^{2}-10H}} $ of the
$\mathcal{PT}$-symmetric $\phi^{6}$ potential.}%
\label{phi6d1}%
\end{figure}

For the $0+1$ case, we get;%
\begin{equation}
\frac{E}{m}=e=-Hb^{6}+\left(  \frac{15}{2t}H+G\right)  b^{4}-\left(  \frac
{45}{4t^{2}}H+\frac{3}{t}G+\frac{1}{2}\right)  b^{2}+\frac{1}{4}%
t+\allowbreak \frac{3}{4t^{2}}G+\frac{15}{8t^{3}}H+\frac{1}{4t},
\end{equation}
where we introduced the dimensionless parameters $b$, $H$, $G$ and $t$ as
$B=m^{\frac{d-2}{2}}\frac{b}{i}$, $h=Hm^{-2d+6}$, $g=Gm^{-d+4}$ and $M=tm$. In
this case we can obtain the dimensionless mass $t=\frac{M}{m}$ of the form;
\[
t=\frac{-15H}{2G-10b^{2}H+\sqrt{4G^{2}+40b^{4}H^{2}-10H}},\allowbreak
\]%
\[
t=\frac{-15H}{2G-10b^{2}H-\sqrt{4G^{2}+40b^{4}H^{2}-10H}}\allowbreak.
\]
As we can see from Fig.\ref{phi6d1}, the effective potential is bounded from
below (stable) for negative $H$ and for $\pm G$ but only for the solution;%

\[
t=\frac{-15H}{2G-10b^{2}H+\sqrt{4G^{2}+40b^{4}H^{2}-10H}}.
\]
In the above calculations, although we used dimensional regularization to
calculate the Feynman diagrams, $E$ is finite even in using direct integral
calculations. In higher dimensions, however, $E$ is divergent even in using
the dimensional regularization and thus one has to follow one of other known
procedures to kill the divergences.

\section{ The effective potential in higher space-time dimensions}

\label{high}

For higher space-time dimensions, the dimensional regularization used to
calculate the Feynman diagrams leaded to the above results may not be able to
get rid of the existing divergences. For instance, in $1+1$ space-time
dimensions, the gamma function, $\Gamma \left(  1-\frac{1}{2}d\right)  $, is
divergent and thus one may resort to another regularization tool like minimal
subtraction. Another point that one has to care about is the invariance of the
bare parameters under the renormalization group. However, as long as we
constrain our calculations up to first order corrections, normal ordering can
overcome these two problems \cite{aboebt,Coleman,normal2,efbook,changcac,mag}.
In the following, we will use the normal ordering technique to study the cases
$1+1$ and $2+1$ while the $3+1$ case will be skipped due to the
non-renormalizability of the theory.

The normal ordering of the field operators follows the relations
\cite{Coleman};
\begin{align*}
N_{m}\psi &  =N_{M}\psi,\\
N_{m}\psi^{2}  &  =N_{M}\psi^{2}+\Delta,\\
N_{m}\psi^{3}  &  =N_{M}\psi^{3}+3\Delta N_{M}\psi,\\
N_{m}\psi^{4}  &  =N_{M}\psi^{4}+6\Delta N_{M}\psi^{2}+3\Delta^{2},\\
N_{m}\psi^{5}  &  =N_{M}\psi^{5}+10\Delta N_{M}\psi^{3}+15\Delta^{2}\psi,\\
N_{m}\psi^{6}  &  =N_{M}\psi^{6}+15\Delta N_{M}\psi^{4}+45\Delta^{2}\psi
^{2}+15\Delta^{3},
\end{align*}
where%
\[
\Delta=\frac{1}{\left(  4\pi \right)  ^{\frac{d+1}{2}}}\left(  \frac
{\Gamma \left(  1-\frac{d+1}{2}\right)  }{\left(  M^{2}\right)  ^{1-\frac
{d+1}{2}}}\right)  ,
\]
and the normal ordering of the kinetic term gives;%
\begin{equation}
N_{m}\left(  \frac{1}{2}\left(  \nabla \psi \right)  ^{2}+\frac{1}{2}\Pi
^{2}\right)  =N_{M}\left(  \frac{1}{2}\left(  \nabla \psi \right)  ^{2}+\frac
{1}{2}\Pi^{2}\right)  +E_{0}(M)-E_{0}(m), \label{norkin1}%
\end{equation}
where
\[
E_{0}(\Omega)=\frac{1}{4}\int \frac{d^{d-1}k}{\left(  2\pi \right)  ^{d-1}%
}\left(  \frac{2k^{2}+\Omega^{2}}{\sqrt{k^{2}+\Omega^{2}}}\right)
=I_{1}+I_{2},
\]
with
\begin{equation}
I_{1}\left(  \Omega \right)  =\frac{1}{2}\frac{1}{\left(  4\pi \right)
^{\frac{d-1}{2}}}\frac{d}{2}\left(  \frac{\Gamma \left(  \frac{1}{2}-\frac
{d-1}{2}-1\right)  }{\Gamma \left(  \frac{1}{2}\right)  }\left(  \frac
{1}{\Omega^{2}}\right)  ^{\frac{1}{2}-\frac{d-1}{2}-1}\right)  ,
\end{equation}%
\begin{equation}
I_{2}\left(  \Omega \right)  =\frac{\Omega^{2}}{4}\frac{1}{\left(  4\pi \right)
^{\frac{d-1}{2}}}\left(  \frac{\Gamma \left(  \frac{1}{2}-\frac{d-1}{2}\right)
}{\Gamma \left(  \frac{1}{2}\right)  }\left(  \frac{1}{\Omega^{2}}\right)
^{\frac{1}{2}-\frac{d-1}{2}}\right)  .
\end{equation}

\begin{figure}[ptb]
\centering
\includegraphics{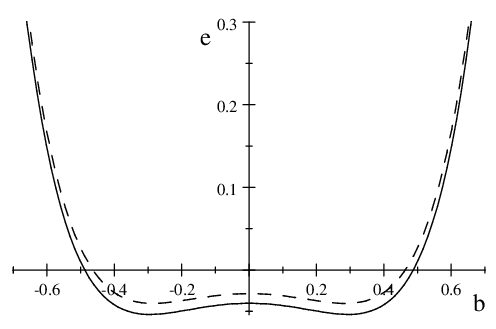} \caption{The $1+1$ space-time dimensions
effective potential as a function of the one-point function measured in units
of $i$ for $H=-\frac{1}{4}\mathtt{,} G=+\frac{1}{2}$ (dashed) and $H=-\frac
{1}{2}\mathtt{,} G=-\frac{1}{2}$ (solid) with the stable solution
$t=\exp \left(  \frac{1}{180H}\left(  -60Hb^{2}+6G+6\sqrt{40H^{2}b^{4}%
+G^{2}-10H}\right)  \right)  $ of the $\mathcal{PT}$-symmetric $\phi^{6}$
potential .}%
\label{phi6d2f}%
\end{figure}

In $1+1$ dimensions, we get $\Delta=-\frac{1}{4\pi}\ln t$ and
\[
E_{0}(M)-E_{0}(m)=\frac{m^{2}}{8\pi}\left(  t-1-\ln t\right)
\]
where $t=\frac{M^{2}}{m^{2}}$. Accordingly, the vacuum energy can take the
form;%
\begin{align*}
\frac{8\pi E}{m^{2}}  &  =e=-2Hb^{6}+\left(  -30H\ln t+G\right)  b^{4}-\left(
90H\ln^{2}t-6\left(  \ln t\right)  G+1\right)  \allowbreak b^{2}\\
&  +3\left(  \ln^{2}t\right)  G-30H\ln^{3}t+t-1-\ln t,
\end{align*}
which is constrained by the equation $\frac{\partial E}{\partial B}=0$ or ;%
\begin{equation}
2b\left(  6Hb^{4}-\left(  2G-60H\ln t\right)  b^{2}+90H\ln^{2}t-6\left(  \ln
t\right)  G+1\right)  \allowbreak=0. \label{2dbeq}%
\end{equation}
In the above forms we used the parametrization $g=2\pi Gm^{2}$, $B=\frac
{b}{i\sqrt{4\pi}}$ and $h=\left(  4\pi \right)  ^{2}Hm^{-2d+6}.$ For $b\neq0$,
Eq.(\ref{2dbeq}) has the solutions,%
\begin{align}
t  &  =\exp \left(  \frac{1}{180H}\left(  -60Hb^{2}+6G+6\sqrt{40H^{2}%
b^{4}+G^{2}-10H}\right)  \right)  ,\nonumber \\
t  &  =\exp \left(  \frac{1}{180H}\left(  -60Hb^{2}+6G-6\sqrt{40H^{2}%
b^{4}+G^{2}-10H}\right)  \right)  .
\end{align}
For the bounded from above case $($negative $h)$, the solution
\[
t=\exp \left(  \frac{1}{180H}\left(  -60Hb^{2}+6G+6\sqrt{40H^{2}b^{4}%
+G^{2}-10H}\right)  \right)  ,
\]
is the stable one (Fig.\ref{phi6d2f}).

\begin{figure}[ptb]
\centering
\includegraphics{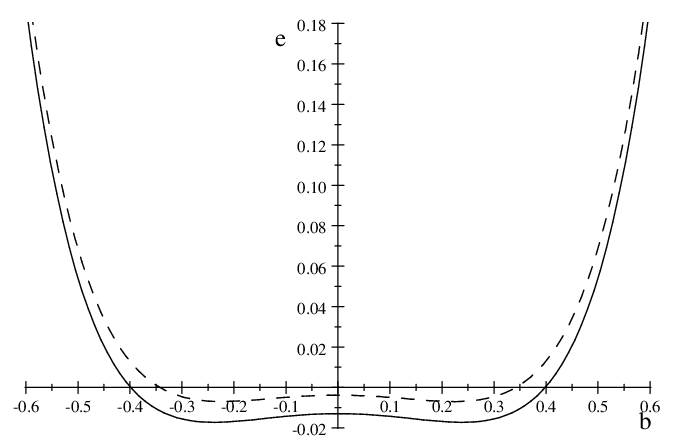}
\caption{The $2+1$ space-time dimensions effective potential as a function of
the one-point function measured in units of $i$ for $H=-1 \mathtt{,}
G=+\frac{1}{2}$ (dashed) and $H=-1{,} G=-\frac{1}{2}$ (solid) with the stable
solution $t=\frac{1}{180H}\left(  180H-60Hb^{2}+12G+6\sqrt{40H^{2}b^{4}%
+4G^{2}-10H}\right)  $ of the $\mathcal{PT}$-symmetric $\phi^{6}$ potential .}%
\label{phi6d3}%
\end{figure}

In $2+1$ dimensions, \ we get $E_{0}(M)-E_{0}(m)=\frac{1}{24\pi}\left(
M^{3}-m^{3}\right)  $ and $\Delta=\frac{1}{4\pi}\frac{m-M}{\pi}$. Using the
parametrization $B=m^{\frac{d-2}{2}}\frac{b}{i\sqrt{\left(  4\pi \right)  }}$,
$g=\left(  4\pi \right)  Gm^{-d+4}$ , $h=\left(  4\pi \right)  ^{2}Hm^{-2d+6}$
and $M=tm$ we obtain the results;%
\begin{align*}
\frac{4\pi E}{m^{3}} &  =e=-Hb^{6}+\left(  15H\left(  1-t\right)  +G\right)
b^{4}-\left(  \frac{1}{2}+45H\left(  t-1\right)  ^{2}+6G\left(  1-t\right)
\right)  \allowbreak b^{2}\\
&  +3G\left(  t-1\right)  ^{2}+15H\left(  1-t\right)  ^{3}+\frac{1}{6}\left(
t+2\right)  \left(  t-1\right)  ^{2},
\end{align*}
and
\begin{align*}
t &  =\frac{1}{180H}\left(  180H-60Hb^{2}+12G+6\sqrt{40H^{2}b^{4}+4G^{2}%
-10H}\right)  ,\\
t &  =\frac{1}{180H}\left(  180H-60Hb^{2}+12G-6\sqrt{40H^{2}b^{4}+4G^{2}%
-10H}\right)  .
\end{align*}
Only the solution
\[
t=\frac{1}{180H}\left(  180H-60Hb^{2}+12G+6\sqrt{40H^{2}b^{4}+4G^{2}%
-10H}\right)  ,
\]
results in a stable effective potential for the bounded from above $\left(
-\phi^{6}\right)  $ \ potential (Fig.\ref{phi6d3}).

\section{comparison of the effective potential predictions with those from the
literature}

\label{comp}

To test our calculations, let us consider the $h=0$ and negative $g$ (
$\mathcal{PT}\emph{-}\phi^{4}$ theory). In this case; Eq.(\ref{veff}) gives;

\bigskip%
\begin{equation}
E=\frac{1}{2}M+\frac{1}{2}B^{2}m^{2}-gB^{4}-\frac{3}{4}\frac{g}{M^{2}}%
+\frac{1}{4M}\left(  -12gB^{2}-M^{2}+m^{2}\right)  .
\end{equation}

Then;
\begin{align}
\frac{\partial E}{\partial B} &  =\frac{\partial}{\partial B}\left(  \frac
{1}{2}M+\frac{1}{2}B^{2}m^{2}-gB^{4}-\frac{3}{4}\frac{g}{M^{2}}+\frac{1}%
{4M}\left(  -12gB^{2}-M^{2}+m^{2}\right)  \right)  \nonumber \\
&  =-\allowbreak4gB^{3}+\frac{1}{M}\left(  Mm^{2}-6g\right)  B
\end{align}

Also, we have the condition%
\begin{align}
\frac{\partial^{2}E}{\partial B^{2}} &  =\frac{\partial^{2}}{\partial B^{2}%
}\left(  \frac{1}{2}M+\frac{1}{2}B^{2}m^{2}-gB^{4}-\frac{3}{4}\frac{g}{M^{2}%
}+\frac{1}{4M}\left(  -12gB^{2}-M^{2}+m^{2}\right)  \right)  \nonumber \\
&  =-12gB^{2}+\frac{1}{M}\left(  Mm^{2}-6g\right)  ,
\end{align}

From Eq.(\ref{EBM}) and for $B\neq0$ \ we get;%
\begin{align}
m^{2}-4gB^{2}-\frac{6g}{M} &  =0\nonumber \\
m^{2}-12gB^{2}-\frac{6g}{M} &  =M^{2},
\end{align}
These are exactly Eq.(39) and Eq.(41)$\allowbreak \allowbreak$ in Ref.
\cite{jonesgr2} where the authors there  obtained them using Schwinger-Dyson equations. Note
that, there, they used the interaction term $-\frac{g}{4}\phi^{4}$ but we used $-g\phi^{4}$
interaction term which should be taken into account in the comparison between
our results and the results in Ref. \cite{jonesgr2}. This result shows that
the algorithm we use is quite reasonable.

\begin{figure}[th]
\begin{minipage}[t]{0.45\linewidth}
\centering
\includegraphics[width=\linewidth]{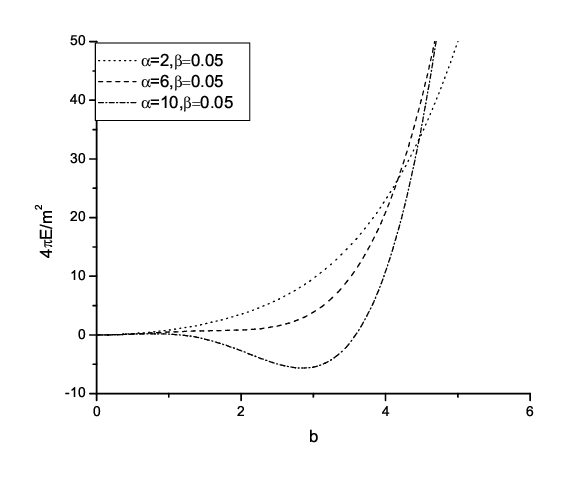}
\caption{In this figure, we plot the effective potential of the Hermitian $(\phi^6)_{1+1}$ theory for positive $\alpha=\frac{G}{4!}$ and fixed $\beta=\frac{H}{4!}$. It is clear that the plot shows a second order phase transition which agrees well with the phase structure of the theory in the literature. }
\label{2nd}
\end{minipage}
\hspace{0.5cm} \begin{minipage}[t]{0.45\linewidth}
\centering
\includegraphics[width=\linewidth]{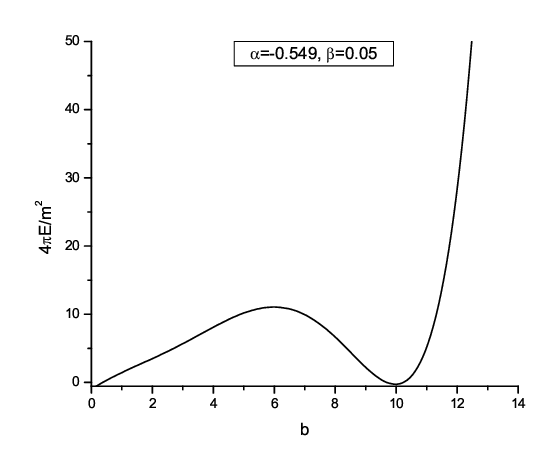}
\caption{In this figure, we plot the effective potential of the Hermitian $(\phi^6)_{1+1}$ theory for negative $\alpha=\frac{G}{4!}$ and fixed $\beta=\frac{H}{4!}$. The plot shows a first order phase transition which agrees with  the predictions  in the literature. }
\label{1st}
\end{minipage}
\end{figure}

Another test to our calculations can be drawn from the phase structure of the
Hermitian $\phi^{6}$ theory (positive $G$ and positive $H$). In fact, we
resort to the Hermitian theory as our calculations for the bounded from above
$\phi^{6}$ theory represents the first study for this theory. For technical
issues,  we use the scaled couplings $\alpha=\frac{G}{4!}$ and $\beta=\frac
{H}{4!}$ to generate the plots in Fig.\ref{2nd} and Fig. \ref{1st}. In these
figures we plot the effective potential for the Hermitian $(\phi^{6})$ field
theory in $1+1$ dimensions. This theory is well known to have a second order
phase transition for positive $G$ while the theory does have a first order
phase transition for negative $G$ \cite{montecarlo}. In fact, our calculations
agrees well with these facts which it represents a good test for the validity
of our calculations.

\section{Conclusions}

\label{conc}

To conclude, we used the canonical quantization method to study non-Hermitian
and $\mathcal{PT}$-symmetric field theory. We showed that the algorithm we
follow can reproduce the same results obtained in the literature (using
Schwinger-Dyson equation) for the $\mathcal{PT} $-symmetric ($-\phi^{4}$)
theory. Also, the algorithm we follow produced the known phase structure for the Hermitian $\phi^6$ field theory. We then extended the algorithm to the study of the $\mathcal{PT}%
$-symmetric ($-\phi^{6}$) theory in $d$ space-time dimensions (for the first
time). Regarding this, we calculated the effective potential of the
$\mathcal{PT}$-symmetric ($-\phi^{6}$) theory in $d$ space-time dimensions.
The classical potential of this theory is bounded from above and thus has not
been stressed in the literature due to the believe that this theory is
instable. We have shown that as long as the vacuum condensate is imaginary,
the effective Hamiltonian is non-Hermitian but $\mathcal{PT}$-symmetric and
the effective potential is rather bounded from below which proves the
stability of the theory. We found three different vacuum solutions however we
figured out that the effective potential is bounded from below for only one
vacuum solution out of the three available solutions. In fact, a lesson can be
learned from this work because it shows that bounded from above potentials can
have both stable and instable vacuum solutions and the bounded from below
potentials can have stable as well as instable solutions too. Accordingly, one
may expect reflections as well as formation of bound states when incident
particles are scattered from either bounded from below or bounded from above
potentials. Predictions of such events in the lab will offer a great support
to the believe in $\mathcal{PT}$-symmetric theories.


\newpage



. \newpage

\newpage

\end{document}